\documentclass{aa}
\usepackage[dvips]{graphicx}

\usepackage{txfonts}

\newcommand{\be}{\begin{equation}}
\newcommand{\ee}{\end{equation}}
\newcommand{\bdm}{\begin{displaymath}}
\newcommand{\edm}{\end{displaymath}}

\begin{document}
   \title{On the mass transfer in AE~Aquarii}
\author{N.R.\,Ikhsanov\inst1\fnmsep\inst3\fnmsep\inst4,
V.V.\,Neustroev\inst2\fnmsep\inst5, and
N.G.\,Beskrovnaya\inst3\fnmsep\inst4}
  \offprints{N.R.~Ikhsanov \\ \email{ikhsanov@kao.re.kr}}

   \institute{Korea Astronomy Observatory, 61-1 Whaam-dong, Yusong-gu,
              Taejon 305-348, Republic of Korea
              \and
              Computational Astrophysics Laboratory, National University
              of Ireland, Galway, Newcastle Rd., Galway, Ireland
              \and
              Central Astronomical Observatory of the Russian
              Academy of Sciences, Pulkovo 65/1, 196140
              St.\,Petersburg, Russia
              \and
              Isaac Newton Institute of Chile, St.Petersburg Branch
              \and
              Isaac Newton Institute of Chile, Kazan Branch
}

   \date{Received 24 February 2004 / accepted 30 March 2004}

\authorrunning{N.R.\,Ikhsanov et~al.}

\abstract{The observed properties of the close binary AE~Aqr
indicate that the mass transfer in this system operates via the
Roche lobe overflow mechanism, but the material transferred from
the normal companion is neither accreted onto the surface of the
white dwarf nor stored in a disk around its magnetosphere. As
previously shown, such a situation can be realized if the white
dwarf operates as a propeller. At the same time, the efficiency of
the propeller action by the white dwarf is insufficient to explain
the rapid braking of the white dwarf, which implies that the
spin-down power is in excess of the bolometric luminosity of the
system. To avoid this problem we have simulated the mass-transfer
process in AE~Aqr assuming that the observed braking of the white
dwarf is governed by a pulsar-like spin-down mechanism. We show
that the expected H$\alpha$ Doppler tomogram in this case
resembles the tomogram observed from the system. We find that the
agreement between the simulated and the observed tomograms is
rather good provided the mean value of the mass-transfer rate
$<\dot{M}> \sim 5\times 10^{16}\,{\rm g\,s^{-1}}$. Three spatially
separated sources of H$\alpha$ emission  can be distinguished
within this approach. The structure of the tomogram depends on the
relative contributions of these sources to the H$\alpha$ emission
and is expected to vary from night to night. \keywords{pulsars --
binaries: close -- magnetic fields -- white dwarfs -- stars:
individual(AE~Aquarii)}}

   \maketitle

   \section{Introduction}

AE~Aquarii is a peculiar nova-like star at a distance of $\sim
100\pm 30$\,pc (Welsh et~al. \cite{whg95}; Friedjung
\cite{friedjung97}). It is a non-eclipsing close binary system
with an orbital period $P_{\rm orb} \approx 9.88$\,hr and orbital
eccentricity $e \approx 0.02$ (Chincarini \& Walker \cite{chw81}).
The normal companion (secondary) is a K3-K5 red dwarf on or close
to the main sequence (Bruch \cite{bruch91}; Welsh et~al.
\cite{whg95}). The primary is a magnetized white dwarf rotating
with the period $P_{\rm s} \approx 33$\,s (Patterson \cite{p79};
Eracleous et~al. \cite{erac94}). The inclination angle of the
system and the mass ratio are limited to $50^{\degr} < i <
70^{\degr}$, and $0.58 \la (q=M_2/M_1) \la 0.89$, respectively,
and the mass of the white dwarf is evaluated as $0.8 \la M_1 \la 1
M_{\sun}$ (Reinsch \& Beuermann \cite{rb94}; Welsh et~al.
\cite{whg95}).

The system emits detectable radiation in almost all parts of the
spectrum. It is a powerful non-thermal flaring radio source
(Bastian et~al. \cite{bdch88}, Meintjes \& Venter \cite{mv03} and
references therein) and, possibly, the source of very high energy
$\gamma$-rays (Bowden et~al. \cite{bbch92}; Meintjes et~al.
\cite{mjr94}; see, however, Lang et~al. \cite{lbc98}). The
optical, UV, and X-ray radiation of the system is predominantly
thermal and comes from at least three different sources. The
visual light is dominated by the secondary (Bruch \cite{bruch91};
Welsh et~al. \cite{whg95}). The contribution of the primary is
observed mainly in the form of 33\,s (and 16.5\,s) coherent
oscillations detectable in the optical, UV, and X-rays (Patterson
\cite{p79}; Eracleous et~al. \cite{erac94}; Patterson et~al.
\cite{pbchr80}). The remaining light comes from a highly variable
extended source, which manifests itself in the blue/UV continuum,
the optical/UV broad single-peaked emission lines, and the
non-pulsing X-ray component. This source is associated with the
mass-transfer process and is suspected of being responsible for
the peculiar rapid flaring of the star (for discussion see e.g.
Eracleous \& Horne \cite{eh96}).

AE~Aqr is currently assigned to the DQ~Her subclass of magnetic
Cataclysmic Variables (CVs). The members of this subclass are
interacting low-mass close binaries, in which the degenerate
companions are magnetized white dwarfs rotating with periods
$P_{\rm s} \ll P_{\rm orb}$ and accreting material from a
Keplerian disk (see e.g. Warner \cite{w95}). However, extensive
investigations during the last decade have clearly shown that
AE~Aqr does not fit in this model. Namely, studies of the 33\,s
pulsations in the optical/UV (Eracleous et~al. \cite{erac94}) and
X-rays (Reinsch et~al. \cite{rbht95}; Clayton \& Osborne
\cite{co95}; Choi et~al. \cite{cda99}) revealed that the
contribution of the white dwarf to the system radiation is
significantly smaller than previously assumed within the
accretion-powered white dwarf model. Furthermore, analysis of the
H$\alpha$ Doppler tomogram of AE~Aqr has shown no evidence of an
accretion disk in the system (Wynn et~al. \cite{wkh97}; Welsh
et~al. \cite{whg98}). Finally, de~Jager et~al. (\cite{jmor94})
reported a mean spin-down rate of the white dwarf $\dot{P}_0 =
5.64 \times 10^{-14}\,{\rm s\,s^{-1}}$. The remarkable stability
of the observed braking over the span of 14.5\,yr suggests that
the entire white dwarf is spinning down (for a detailed discussion
see Welsh \cite{welsh99}). This allows us to evaluate the
spin-down power of the white dwarf as
   \be
L_{\rm sd} = 6 \times 10^{33}\ I_{50}\ P_{33}^{-3}\
(\dot{P}/\dot{P}_0)\ {\rm erg\,s^{-1}},
  \ee
where $I_{50}$ and $P_{33}$ are the moment of inertia and the spin
period of the white dwarf expressed in units of $10^{50}\,{\rm
g\,cm^2}$ and 33\,s, respectively. $L_{\rm sd}$ exceeds the
luminosity of the system observed in the UV and X-rays by a factor
of 120--300 and even the bolometric luminosity by a factor of more
than 5 (hereafter, the distance to the star is adopted as
100\,pc). This indicates that the spin-down power dominates the
system energy budget and raises a question about the form in which
this energy is released.

Among possible answers to this question the following two are
currently under discussion. The first, a so called `magnetic
propeller' model, was presented by Wynn et~al. (\cite{wkh97}), who
suggested that the rotation rate of the white dwarf decelerates by
means of interaction between its fast rotating magnetosphere and
the material inflowing from the secondary. The second, a so called
`pulsar-like white dwarf' model, was suggested by Ikhsanov
(\cite{i98}), who indicated that the observed braking of the white
dwarf could be explained in terms of the canonical pulsar-like
spin-down mechanism (Pacini \cite{pac68}; Goldreich \& Julian
\cite{gj69}), provided its surface magnetic field is 50\,MG. In
this paper we address the comparative analysis of these models.
The basic statements of these approaches are briefly discussed in
the following two sections. In Sect.\,4 we present the results of
our simulation of the H$\alpha$ Doppler tomogram of AE~Aqr. The
adopted assumptions are summarized in Sect.\,5. The basic
conclusions are given in Sect.\,6.

  \section{Propeller action by the white dwarf}

 The observed properties of the optical/UV emission lines
unambiguously indicate that a relatively intensive mass-transfer
takes place between the system components of AE~Aqr. In
particular, the narrow component of the Balmer emission lines is
nearly in anti-phase to the absorption lines of the red dwarf
(Reinsch \& Beuermann \cite{rb94}), that suggests its origin is
near the white dwarf. Furthermore, the evaluated velocity and
luminosity of the emission line source are significantly larger
than those typically expected in the wind of red dwarfs (see e.g.
Eracleous et~al. \cite{erac94}).These properties speak in favor of
an association of the emission line source with the material
transferred from the normal component through the Roche lobe of
the white dwarf.

The rate of mass transfer in AE~Aqr is still a subject of
discussion. A lower limit to this parameter can be derived
assuming that the radiation of emission lines is generated inside
the Roche lobe of the white dwarf and is powered by the accretion
energy. In this case one finds $\dot{M} > 10^{15} L_{31} R_{10}
M_{0.9}^{-1}\,{\rm g\,s^{-1}}$, where $L_{31}$ is the luminosity
of the emission line source (see e.g. Table~3 in Eracleous \&
Horne \cite{eh96}), and $M_{0.9}$ is the mass of the white dwarf
in units of $0.9\,M_{\sun}$. $R_{10}$ is the distance of the
closest approach of the material responsible for the observed
emission lines to the white dwarf expressed in units of
$10^{10}$\,cm. This parameter can be limited using the expression
$R\la GM_{\rm wd}/V_{\rm em}^2$, where $V_{\rm em}$ is the
velocity of the emitting material derived from the observed width
of the emission lines.

The above limit to $\dot{M}$ represents the minimum possible value
of the mass-transfer rate in AE~Aqr and, as will be shown below,
is significantly underestimated. Nevertheless, this estimate plays
an important role in the identification of the mass-transfer
mechanism. Indeed, the derived value exceeds the maximum possible
rate of mass capture by the white dwarf from the wind of its
companion by more than three orders of magnitude (Ikhsanov
\cite{i97}). This justifies that the mass-transfer in AE~Aqr
operates via the Roche lobe overflow mechanism and hence, the
secondary overflows its Roche lobe and loses material through the
L1 point towards the primary.

However, a relatively low X-ray luminosity of AE~Aqr ($L_{\rm x}
\sim 10^{31}\,{\rm erg\,s^{-1}}$, see e.g. Choi et~al.
\cite{cda99}) and the structure of the H$\alpha$ Doppler tomogram
derived by Wynn et~al. (\cite{wkh97}) and Welsh et~al.
(\cite{whg98}) argue against the possibility that the material
transferred from the red dwarf is either accreted onto the surface
of the white dwarf or stored in a disk. To solve this paradox the
hypothesis has been invoked that the material flowing into the
Roche lobe of the white dwarf is ejected from the system without
forming a disk.

An effort to reconstruct the mass-transfer picture within this
hypothesis was first made by Wynn et~al. (\cite{wkh97}). They
modelled the stream as a set of diamagnetic blobs, which move
through the fast rotating magnetosphere of the primary,
interacting with the local magnetic field via a surface drag term.
In this case, the trajectories of the blobs differ from the
ballistic case, and the stream is able to leave the system without
forming a disk if the magnetic moment of the primary is $\mu \ga
10^{32}\,{\rm G\,cm^3}$. In particular, putting $\mu \simeq
10^{32}\,{\rm G\,cm^3}$ (i.e. within the expected range of the
magnetic moments of Intermediate Polars), Wynn et~al.
(\cite{wkh97}) found that blobs reach the escape (maximum)
velocity of $V_{\rm esc} \la 1000\,{\rm km\,s^{-1}}$ at the
closest approach to the white dwarf, $r_0\ga 10^{10}$\,cm, and
leave the system without forming a disk with an average velocity
$V_{\infty} \sim 300\,{\rm km\,s^{-1}}$. The ejection of material
in this scenario occurs due to propeller action by the white
dwarf, which is also assumed to be responsible for the observed
braking of the primary.

The H$\alpha$ Doppler tomogram calculated within this model is
similar to the tomogram observed in AE~Aqr in several important
aspects. In particular, neither shows azimuthal symmetry, and the
emission is not centered on the white dwarf but is primarily in
the lower-left quadrant ($V_{\rm x}$, $V_{\rm y}$ both negative).
These similarities indicate that the picture reconstructed by Wynn
et~al. (\cite{wkh97}) is qualitatively correct.

 At the same time, some of the quantitative predictions
of the `magnetic propeller' model have not been observationally
confirmed. As shown by Welsh et~al. (\cite{whg98}), the observed
tomogram does not contain the high velocity `loop' ($|V| \sim
700-1000\,{\rm km\,s^{-1}}$) predicted by Wynn et~al.
(\cite{wkh97}, see Fig.\,3), and on the other hand it shows that
the contribution of material at low velocities ($|V| \la 100\,{\rm
km\,s^{-1}}$) is significantly larger than expected from the
simulated picture. These inconsistencies forced Welsh et~al.
(\cite{whg98}) to suggest that the heating of the blobs, when they
pass the acceleration region at their closest approach to the
white dwarf, is negligible, and therefore that their contribution
to the H$\alpha$ emission of the system is small. Following this
assumption they have placed the region of energy release outside
the Roche lobe of the primary where the trajectories of blobs of
different masses cross each other and collisions of the ejected
blobs can be expected. However, as mentioned by Welsh
(\cite{welsh99}), some of the properties of AE~Aqr (such as the
large velocities in the emission lines during flares and the
existence of high-excitation emission lines) remain puzzling even
in this so called `colliding blobs' scenario.

Another difficulty with the `magnetic propeller' model has been
mentioned by Ikhsanov (\cite{i98}), Meintjes \& de\,Jager
(\cite{mdj00}), and Choi \& Yi (\cite{cy00}). As they have shown,
the efficiency of the propeller action by the white dwarf under
the conditions of interest is not sufficient to explain the
observed rapid braking of the primary. Indeed, following Wynn
et~al. (\cite{wkh97}) one could assume that almost all spin-down
power of the white dwarf is transferred into the kinetic energy of
the ejected material. However, the kinetic luminosity of the
ejected blobs is obviously limited to
  \be
L_{\rm kin} \la (1/2) \dot{M} V_{\rm esc}^2 \simeq 5 \times
10^{32}\dot{M}_{17}\ V_8^2\ {\rm erg\,s^{-1}},
  \ee
where $V_8=V_{\rm esc}/10^8\,{\rm cm\,s^{-1}}$ and $\dot{M}_{17}$
is the mass transfer rate expressed in units of $10^{17}\,{\rm
g\,s^{-1}}$. Hence, for this assumption to be satisfied, the mass
transfer rate in the system should be in excess of $10^{18}\,{\rm
g\,s^{-1}}$, which is inconsistent with the value of $\dot{M}$
derived from observations (see e.g. Eracleous \& Horne
\cite{eh96}). On the other hand, the assumption that a significant
part of the spin-down power is transferred into the thermal energy
of the ejected gas contradicts the derived value of the ratios
$L_{\rm UV}/L_{\rm sd} \sim L_{\rm x}/L_{\rm sd} \ll 1$.
Therefore, the question about the nature of the observed braking
of the white dwarf within the `magnetic propeller' model remains
open.

The problems mentioned above indicate that the `magnetic
propeller' model cannot provide us with a complete picture of
AE~Aqr, and an improvement of this model is required. They also
show that the problem that should be addressed first in any
further improvements is the spin-down mechanism of the white
dwarf. As long as this problem remains unsolved, the form in which
the spin-down power is released turns out to be unclear, and
therefore, the major part of the energy released in the system is
not taken into account.

At the same time, for a solution of the spin-down problem to be
reliable it should also meet the diskless mass-transfer criteria.
In this light, the improvement suggested by Meintjes \& de\,Jager
(\cite{mdj00}) cannot be applied to the interpretation of AE~Aqr,
since their approach requires the existence of a clumpy disk
around the white dwarf. On the other hand, the model of Choi \& Yi
(\cite{cy00}), in which the spin-down power is assumed to be spent
in the emission of gravitational waves, cannot be accepted either.
Although the mass-transfer picture within this model is similar to
that reconstructed by Wynn et~al. (\cite{wkh97}), the adopted mass
distribution over the primary surface is unreliable (for a
detailed discussion see Ikhsanov \& Beskrovnaya \cite{ib02}).
Among the improvements of the `magnetic propeller' model so far
discussed in the literature, only the pulsar-like spin-down model
meets the criteria. The reliability of this improvement is
discussed in the following section.

   \section{Pulsar-like spin-down}

The hypothesis of pulsar-like spin-down of the white dwarf in
AE~Aqr has a certain observational basis. A situation in which the
spin-down power of a star exceeds its bolometric luminosity
significantly is unique for CVs as well as for all presently known
accretion-powered sources. At the same time, this situation is
typical for the spin-powered pulsars, whose luminosity constitutes
only a small fraction ($\sim 10^{-3} - 10^{-1}$) of their
spin-down power (for a review see Manchester \& Taylor
\cite{mt77}; Hartmann \cite{har95}). Furthermore, while the
appearance of AE~Aqr in X-rays is very atypical for the
accretion-powered white dwarfs (Clayton \& Osborne \cite{co95}),
it resembles the appearance of spin-powered pulsars observed in
the {\sc ROSAT} energy range (see e.g. Becker \& Tr\"umper
\cite{bt97}). For instance, the X-ray spectrum is significantly
softer than those typically observed from accretion-powered
compact stars, and the ratio of the luminosity of the pulsing
component to the spin-down power is close to $10^{-3}$. Finally,
as reported by Meintjes et~al. (\cite{mjr94}), the intensity of
the very high energy $\gamma$-ray emission detected from AE~Aqr
significantly exceeds the intensity of radiation emitted in other
parts of the spectrum. Such behavior is also typical for
spin-powered pulsars (see e.g. Tompson \cite{tom96}) and is
consistent with modern views on the processes of energy release in
these sources (for a review see Michel \cite{m91}). Thus, the
investigation of the hypothesis that the braking of both the white
dwarf in AE~Aqr and the spin-powered pulsars is governed by the
same mechanism appears to be quite reasonable.

As shown by Ikhsanov (\cite{i98}), for this hypothesis to be
effective the dipole magnetic moment of the white dwarf should be
as large as
   \be\label{magmom}
\mu \simeq 1.4 \times 10^{34}\ P_{33}^{2} \left(\frac{L_{\rm
sd}}{6 \times 10^{33}\,{\rm erg\,s^{-1}}}\right)^{1/2} {\rm
G\,cm^{3}}.
   \ee
This implies that the mean strength of the magnetic field at the
surface of the white dwarf is $B(R_{\rm wd}) \approx 50$\,MG.
Under these conditions the spin-down power is spent in the
generation of magneto-dipole waves and particle acceleration and
hence is released mainly in undetectable parts of the spectrum.
Therefore, the observed inequality $L_{\rm bol} < L_{\rm sd}$
turns out to be naturally explained within this approach.

The limitation of the magnetic field to 50\,MG is consistent with
present views on possible values of the surface field strength of
white dwarfs (see e.g. Jordan \cite{jord01}). In particular, the
magnetic field of white dwarfs in Polars is of the same order of
magnitude (Cropper \cite{crop90};  Chanmugam \cite{chanm92}).
However, it is significantly above the previous limit to the
strength of the magnetic field of the white dwarf in AE~Aqr
derived by Bastian et~al. (\cite{bdch88}) and Stockman et~al.
(\cite{ssb92}) from the analysis of the circularly polarized
optical emission.

The reason for this inconsistency has recently been investigated
by Ikhsanov et~al. (\cite{ijb02}). As they have shown, the
limitation presented by Bastian et~al. (\cite{bdch88}) and
Stockman et~al. (\cite{ssb92}) is model-dependent and is based on
the assumption that the radiation of the white dwarf is powered
mainly by the accretion of material onto its surface. However, the
investigations of AE~Aqr in the UV (Eracleous et~al.
\cite{erac94}) and X-rays (Clayton \& Osborne \cite{co95}; Choi
et~al. \cite{cda99}) have clearly shown that this assumption is
not valid. As presently recognized, the contribution of the hot
polar caps to the visual radiation of AE~Aqr does not exceed
0.1\%--0.2\%. In this situation the hot polar caps cannot be the
source responsible for the circularly polarized radiation observed
from the system. Otherwise, the intrinsic polarization of the
source proves to be in excess of 100\%, that is obviously
impossible (see Ikhsanov et~al. \cite{ijb02}). Therefore, the
above mentioned inconsistency cannot be used as an argument to
reject the possibility of the white dwarf in AE~Aqr having a
magnetic field as strong as 50\,MG.

A possible history of AE~Aqr within the pulsar-like model is more
complicated than that usually modelled within the `magnetic
propeller' approach (Meintjes \cite{m02}; Schenker et~al.
\cite{skkwz02}). Indeed, a white dwarf with mass $0.9\,M_{\sun}$
and surface magnetic field 50\,MG can only be spun up to the
period of 33\,s if the mass transfer rate during a previous epoch
was in excess of the Eddington limit ($\sim 3\times 10^{21}\,{\rm
g\,s^{-1}}$) by a factor of 3. Accretion with these
characteristics resembles the process of the merging of a white
dwarf with another star rather than the mass exchange between a
main sequence red dwarf and a white dwarf in a close binary. The
formation of a fast rotating, strongly magnetized white dwarf due
to its merging with a companion has been already discussed by
Paczy\'nski (\cite{pac90}). Following this scenario, one should
assume that the white dwarf in AE~Aqr is a product of the merging
of a magnetized white dwarf and, possibly, a brown dwarf of mass
$\sim 0.03 M_{\sun}$.

However, as pointed out by Ikhsanov (\cite{i99}), the process of
merging is not the only possible solution. An alternative
explanation is based on the scenario of magnetic field
amplification in very fast rotating compact stars ( Klu\'zniak \&
Ruderman \cite{kr98}; Spruit \cite{s99}). According to Chanmugam
et~al. (\cite{cmt87}), the rotation of a white dwarf with mass
$M_{\rm wd}=0.9\,M_{\sun}$ becomes significantly non-uniform as
its period decreases below $P_{\rm cr} \simeq 20$\,s. The magnetic
field inside the star in this state is winding up to $\sim
10^9$\,G on a time scale of a month, and manifests itself at the
surface due to the buoyant instability producing a surface field
of $\sim 10^8$\,G. This allows us to envisage a situation in which
the magnetic field of the white dwarf in AE~Aqr was amplified to
its present value during the last stage of a previous
accretion-driven spin-up epoch. As shown by Ikhsanov (\cite{i99}),
for this scenario to be effective one has to assume that the
initial magnetic moment of the white dwarf was
  \bdm
\mu_0 \la 2 \times 10^{31}\ \dot{M}_{17}^{1/2}\ P_{20}^{7/6}\
M_{0.9}^{5/6}\ {\rm g\,s^{-1}},
    \edm
which implies that in a previous epoch AE~Aqr was an ordinary
member of the DQ~Her subclass of CVs.

Although both of the above mentioned scenarios lead to a rather
complicated history of AE~Aqr, it is clear that a solution of this
problem within the pulsar-like spin-down model is not impossible.
A more precise investigation, however, is not effective as long as
the presently observed stage of the system is not well identified.
Therefore, in this paper we will focus mainly on the analysis of
currently observed properties of AE~Aqr.

The natural solution of the spin-down problem is not the only
advantage of the pulsar-like model. It also gives a reasonable
explanation of some properties of AE~Aqr observed in the
high-energy parts of the spectrum. In particular, it predicts the
maximum energy of particles accelerated by the white dwarf $E_{\rm
p}^{\rm max} \sim 2\times 10^{12}$\,eV (see Eq.~10 in Ikhsanov
\cite{i98}). This prediction is consistent with the
characteristics of the TeV $\gamma$-ray events observed from the
system (Meintjes et~al. \cite{mjr94}). Furthermore, it also allows
us to associate the origin of the pulsing UV and X-ray emission
with the processes of non-thermal energy release in the
magnetosphere of the white dwarf, i.e. particle acceleration in
the inner (and, possibly, outer) gap and, correspondingly, the
impact of particles responsible for the back-flowing current onto
the surface of the white dwarf at the magnetic pole regions (for a
discussion see Ikhsanov \cite{i98} and references therein). The
latter process should lead to the heating of the surface of the
white dwarf, while the radiative losses of relativistic electrons
are expected to be observed in the X-ray part of the spectrum.
Under these conditions, the luminosity of pulsing emission in the
UV would be comparable with that of the pulsing X-ray component,
and the area of the hot polar caps can be limited to (see Eq.~11
in Ikhsanov \cite{i98})
  \be
A_{\rm pc} \la 3 \times 10^{16}\ \eta_{0.37}^{-1}\
\dot{M}_{16.5}^{2/7}\ \mu_{34.2}^{-4/7}\ R_{8.8}^3\ M_{0.9}\ {\rm
cm^2},
   \ee
where $\mu_{34.2}$ and $R_{8.8}$ are the magnetic moment and the
radius of the white dwarf expressed in units of $10^{34.2}\,{\rm
G\,cm^3}$ and $10^{8.8}$\,cm, respectively.
$\eta_{0.37}=\eta/0.37$ is the parameter accounting for the
geometry of the accretion flow, which in the case of a stream is
normalized following Hameury et~al. (\cite{hkl86}).

Both of these predictions are in good agreement with corresponding
properties of AE~Aqr (see e.g. Eracleous et~al. \cite{erac94};
Choi et~al. \cite{cda99}), and they allow us to avoid a very
controversial assumption about the accretion nature of the pulsing
UV and X-ray radiation of AE~Aqr (for a discussion see Choi et~al.
\cite{cda99} and Ikhsanov \cite{i01}).

Finally, the pulsar-like model naturally leads to a conclusion
about the diskless mass transfer in the system. Indeed, within
this approach the Alfv\'en radius of the white dwarf,
  \be\label{ar}
R_{\rm A} \simeq 3 \times 10^{10}\ \eta_{0.37}\ \mu_{34.2}^{4/7}\
\dot{M}_{17}^{-2/7}\ M_{0.9}^{-1/7}\ {\rm cm},
 \ee
is larger than the circularization radius for all reasonable
values of $\dot{M}$. This means that diskless mass transfer in the
system is realized, and moreover, this is expected independently
of whether the stream transferred through the L1 point is
initially homogeneous or inhomogeneous. In this case the
assumption that the stream disintegrates into blobs before the
point of its closest approach to the white dwarf (as adopted by
Wynn et~al. \cite{wkh97}) turns out to be unnecessary. Instead,
one can envisage a scenario in which the stream disintegration
occurs in the vicinity of the Alfv\'en surface of the white dwarf,
where such a disintegration, according to Arons \& Lea
(\cite{al80}), is expected. However, is the H$\alpha$ Doppler
tomogram expected within this scenario consistent with the
tomogram observed from the system\,? The analysis of this question
is addressed in the next section.

    \section{Simulation of H$\alpha$ Doppler tomogram}

We consider a situation in which the secondary star overflows its
Roche lobe and loses material in the form of a stream through the
L1 point. The stream flows into the Roche lobe of the white dwarf
at the speed of sound and initially follows a ballistic
trajectory. Following Wynn et~al. (\cite{wkh97}), we assume that
at a certain point (its location depends on the scenario
considered) the stream disintegrates into a set of large
diamagnetic blobs. The blobs interact with the magnetospheric
field of the white dwarf via the drag term. Due to this
interaction their trajectories are modified by the magnetic
acceleration
 \be
g_{\rm mag} = -k[V-V_{\rm f}]_{\bot},
  \ee
where $k \sim B^2/c_{\rm A}\rho_{\rm b} l_{\rm b}$ is the drag
coefficient, $V$ and $V_{\rm f}$ are the blob and field
velocities, and the suffix $\bot$ denotes the velocity component
perpendicular to the field lines. $c_{\rm A}$ is the Alfv\'en
speed in the interblob plasma, which under the conditions of
interest can be approximated by the speed of light. $\rho_{\rm b}$
and $l_{\rm b}$ are the density and radius of the blobs.

As shown by Wynn \& King (\cite{wk95}), the drag coefficient can
be expressed in the form $k \sim k_0(r/r_0)^{-n}$, where $k_0$,
$n$, and $r_0$ are constants. Setting $\rho_{\rm b}(r_0)=10^{-11}
\rho_{-11}\,{\rm g\,cm^{-3}}$ and $l_{\rm b}(r_0)=10^9 l_9$\,cm we
evaluate the parameter $k_0$ as (for discussion see Wynn et~al.
\cite{wkh97})
 \be
k_0 \simeq 3.3 \times 10^{-9}\ B^2(r_0)\ \rho_{-11}^{-1}\
l_9^{-1}\ {\rm s^{-1}}.
 \ee

Simulating the trajectory of the material within the Roche lobe of
the white dwarf we have assumed that the drag interaction between
the stream and the magnetic field of the primary before the point
of the stream disintegration is small and therefore, the stream
before this point follows a ballistic trajectory. This assumption
is reasonable if the point of the stream disintegration is located
at a distance $r \ga R_{\rm circ}$, where $R_{\rm circ}$ is the
circularization radius, which is about $2.5 \times 10^{10}$\,cm
for the parameters of AE~Aqr (see Eq.~4.17 in Frank et~al.
\cite{fkr85}). Starting with the point at which the stream
disintegrates into blobs the drag term becomes important and
therefore, the trajectories of the blobs deviate significantly
from the ballistic ones. Following this method, we have performed
several runs of calculations placing the point of disintegration
at the L1 point, at the point of the closest approach of the
stream to the white dwarf, $r_0$, and at a few intermediate points
located within the interval [L1, $r_0$]. This scheme of
calculations has been applied to both the `magnetic propeller' and
the `pulsar-like white dwarf' approaches. The results of our
simulations are presented in the following two subsections.

  \subsection{The magnetic propeller approach}


\begin{figure*}
  \includegraphics[width=16cm]{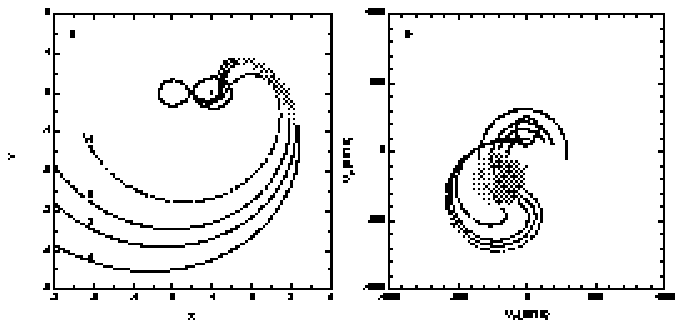}
  \includegraphics[width=8.0cm]{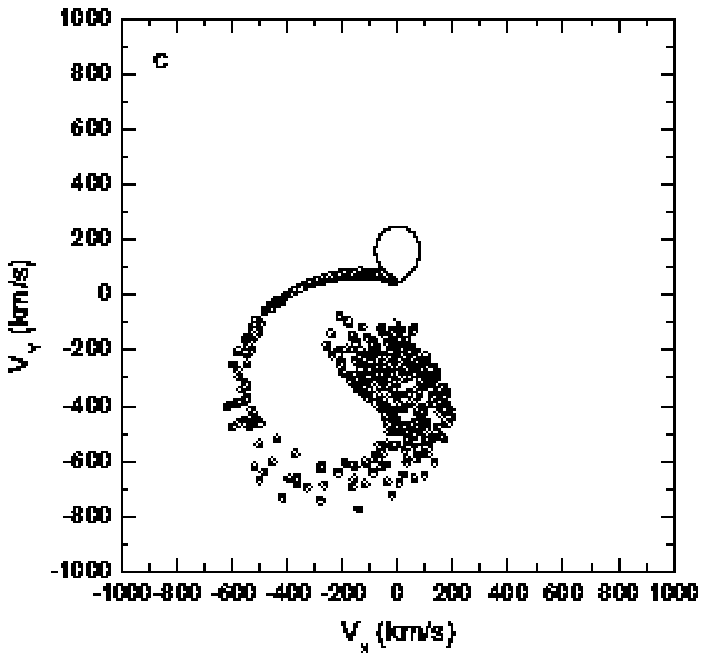}
  \caption{
   Trajectories of blobs in coordinate {\bf a} and
  the velocity {\bf b} and {\bf c} space  within the ``magnetic propeller'' approach.
  The position of the white dwarf is marked by the cross. The
  Roche lobe is shown for both system components in panel\,{\bf a},
  and for the normal component in panels\,{\bf b} and {\bf c}. The system separation
  is used as a unit scale in panel\,{\bf a}.
  The lines 1-4 represent the trajectories of blobs of different
  masses in descending order with density.
  The hatched region in panels {\bf a} and {\bf b} indicates the
  location of the collisions zone of the blobs (see text for further details).}
  \label{cbtom}
\end{figure*}


To test the consistency of the results of our calculations with
those previously derived by other authors we have simulated the
trajectories of blobs using the parameters of AE~Aqr adopted by
Wynn et~al. (\cite{wkh97}) as follows. System parameters: mass
ratio $q=0.64$, orbital period $P_{\rm orb}=9.9$\,hr, and
inclination angle $i=55^{\degr}$, and parameters of the white
dwarf: mass $M_1=0.9 M_{\sun}$, dipole magnetic moment
$\mu=10^{32}\,{\rm G\,cm^3}$, and spin period $P_{\rm s}=33$\,s.
The stream of material transferred from the secondary has been
modelled as a set of diamagnetic blobs. The distance to the point
of the stream disintegration has been assumed to satisfy the
condition $r_{\rm dis} \gg R_{\rm circ}$. The radius and the
density of blobs at their closest approach to the white dwarf have
been taken as $l_{\rm b}=10^9\,l_9$\,cm and $\rho_{\rm
b}=10^{-11}\,\rho_{-11}\,{\rm g\,cm^{-3}}$, respectively. The
value of the parameter $n$ has been chosen to be $n=2$. Finally,
the parameter $r_0$ has been limited to $r_0 \ga R_{\rm min}$,
where
  \be\label{rmin}
 R_{\rm min} \simeq 10^{10} \left(\frac{q}{0.64}\right)^{ - 0.464}
 \left(\frac{a}{1.8 \times 10^{11}\,{\rm cm}}\right)\ {\rm cm}
   \ee
is the distance of the closest approach of a homogeneous stream to
the white dwarf (see Eq.~2.14 of Warner \cite{w95}).

The trajectories of the blobs and the expected H$\alpha$ Doppler
tomogram simulated under these conditions are shown in
Fig.\,\ref{cbtom}. The best fit to the observed tomogram is found
for $k_0 \simeq (0.8-1.3) \times 10^{-5}\,{\rm s^{-1}}$. Lines
1--4 represent the trajectories of blobs of different mass with
the mass of the blobs decreasing from line~1 to line~4. The more
massive the blob the smaller the distance to which it approaches
the white dwarf. The trajectories of blobs of different masses
intersect beyond the Roche lobe of the white dwarf and, therefore,
collisions of the ejected blobs in this region can be expected.
The location of the collision zone is shown in panels~a and b of
Fig.\,\ref{cbtom} as a hatched region. This region represents the
expected structure of the H$\alpha$ Doppler tomogram within the
'colliding blobs' model. The structure of the tomogram derived
within the approach of Wynn et~al. (\cite{wkh97}), i.e. under the
assumption that the H$\alpha$ emission comes mainly from blobs
moving through the Roche lobe of the white dwarf, is shown in
panel~c.

As is easy to see, the derived tomograms are almost identical to
those presented by Wynn et~al. (\cite{wkh97}, see Fig.\,3 of their
paper) and Welsh et~al. (\cite{whg98}, see Fig.\,14 of their
paper). This proves that our code is working properly and can be
used for further analysis.

 \subsection{The pulsar-like white dwarf approach}

\begin{figure*}
 \vspace*{1cm}
 \centerline{\resizebox{12cm}{!}{\includegraphics{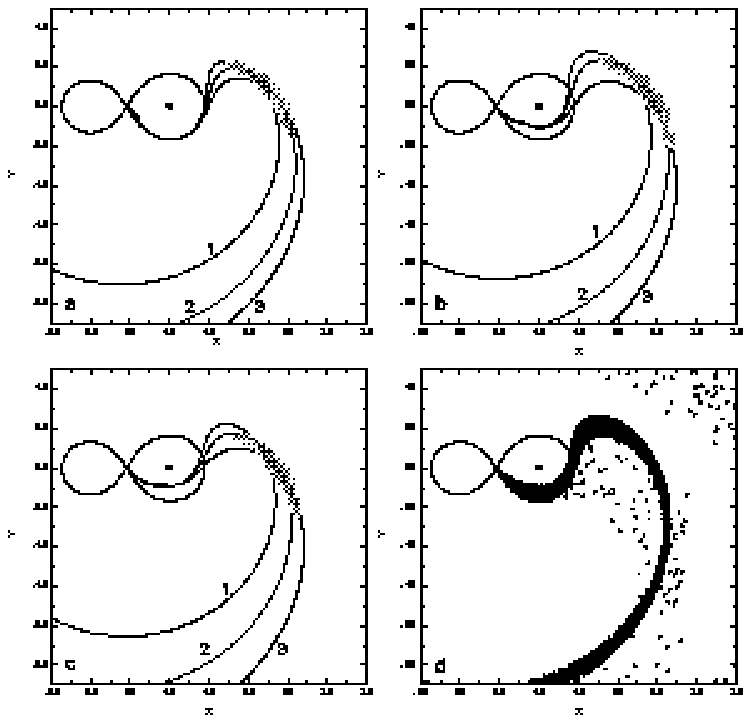}}}
 \vspace*{10mm}
 \centerline{\resizebox{10cm}{!}{\includegraphics{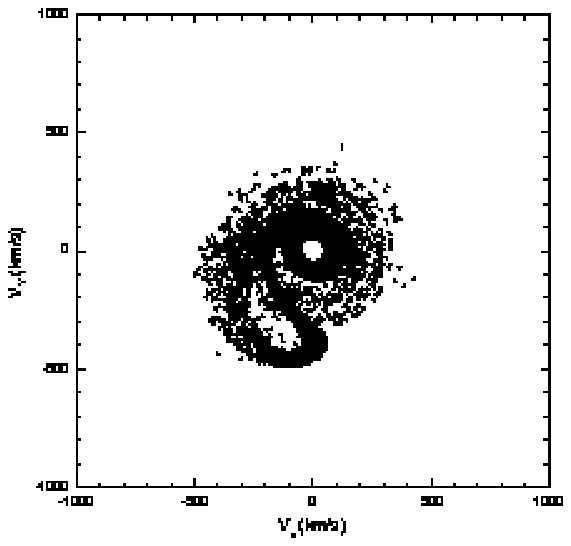}}}
 \caption{   Trajectories of blobs in the coordinate ({\it upper panels} {\bf a-d})
 and the velocity ({\it bottom panel}) space within the pulsar-like model of AE~Aqr.
 The position of the white dwarf is marked with a black dot.
 Panels\,{\bf a}, {\bf b}, and {\bf c} were calculated for mass transfer rates
 of  $10^{16}\,{\rm g\,s^{-1}}$, $5\times 10^{16}\,{\rm g\,s^{-1}}$, and
 $10^{17}\,{\rm g\,s^{-1}}$, respectively. The lines in these panels show the
 trajectories of blobs of different masses. Panel\,{\bf d} is the
 superposition of panels {\bf a}, {\bf b}, and {\bf c} taken with similar
 weights. The bottom panel represents the expected structure of the H$\alpha$ Doppler
 tomogram of the system within this approach. The calculations were made
 on the assumption that the stream is disintegrated into the blobs at the L1 point.
 For a detailed description see text.}
 \label{tr1}
\end{figure*}

\begin{figure*}
 \vspace*{1.0cm}
 \centerline{\resizebox{12cm}{!}{\includegraphics{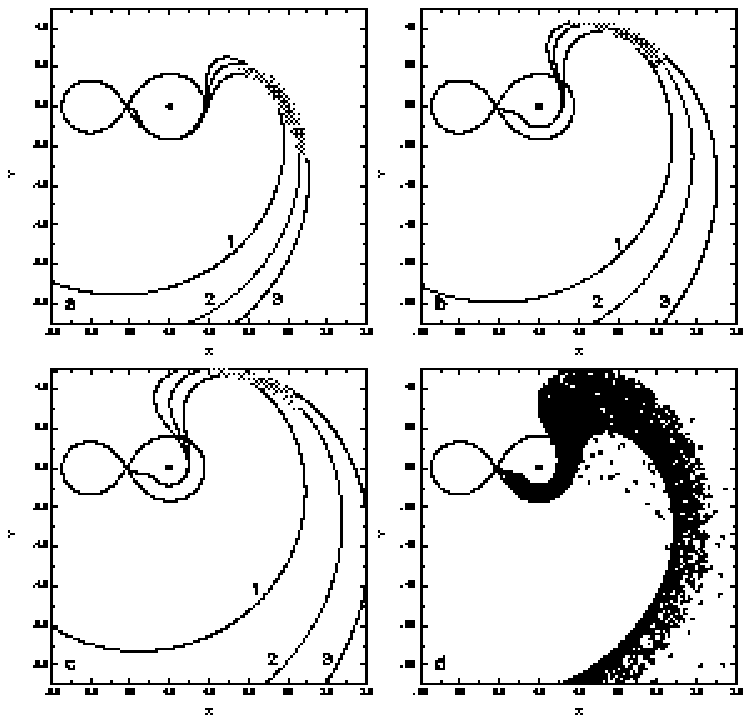}}}
 \vspace*{1.0cm}
 \centerline{\resizebox{10cm}{!}{\includegraphics{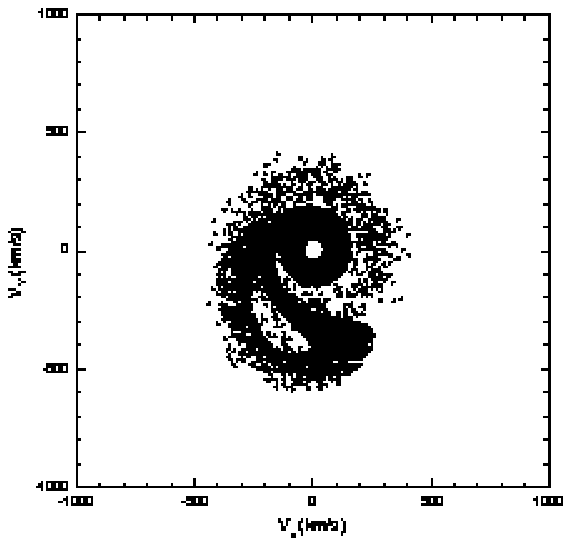}}}
 \caption{The same as Fig\,\ref{tr1}, except for the assumption that
 the point of the stream disintegration is located at the Alfv\'en
 surface of the white dwarf.}
 \label{tr2}
\end{figure*}

The simulation of the stream trajectory within the `pulsar-like`
model differs from that in the frame of the `magnetic propeller'
model in several important aspects. First, the dipole magnetic
moment of the white dwarf within the `pulsar-like' model is
assumed to be $\mu \simeq 1.4 \times 10^{34}\,{\rm G\,cm^3}$, i.e.
a factor of 100 larger than that adopted within the `magnetic
propeller' model.

Second, the value of the parameter $r_0$ is limited to
  \be\label{ralf}
r_0 \ga R_{\rm A} \simeq 3 \times 10^{10}\ \eta_{0.37}\
\mu_{34.2}^{4/7}\ \dot{M}_{17}^{-2/7}\ M_{0.9}^{-1/7}\ {\rm cm}.
 \ee
This limitation reflects the fact that the magnetic field pressure
at the Alfv\'en radius is strong enough to prevent the stream from
approaching the white dwarf closer than $R_{\rm A}$. Within the
`magnetic propeller' model $R_{\rm A} < R_{\rm min}$, and hence
the pressure by the dipole field of the primary in the radial
direction (within the frame centered at the white dwarf) can be
neglected. But in the case of the `pulsar-like' model the Alfv\'en
radius of the white dwarf is larger than $R_{\rm min}$. This means
that the radial velocity of the inflowing material at $R_{\rm A}$
rapidly drops to zero with corresponding heating and possibly
disintegration of the initial stream. The further trajectory of
the material depends on the efficiency of the drag interaction
between the blobs and the magnetospheric field of the primary. If
this interaction is effective enough for the blobs to be
accelerated to the escape velocity in the azimuthal direction,
they will leave the system. Otherwise, the material will follow an
almost circular orbit around the magnetosphere of the white dwarf.

Finally, the assumption about the stream disintegration at the L1
point within the `pulsar-like' model is not necessary. The point
at which the stream disintegrates into diamagnetic blobs in this
case can be located anywhere between the L1 point and the Alfv\'en
surface of the primary. As mentioned above, in both cases
(homogeneous and inhomogeneous stream) diskless mass transfer is
expected within this model. Therefore, parallel to the
traditionally considered case of an inhomogeneous stream at the L1
point we also simulated the mass transfer assuming that the point
of disintegration of the initially homogeneous stream is located
at the Alfv\'en surface of the primary.

Evaluating the structure of the H$\alpha$ Doppler tomogram, one
should also take into account that a third source of H$\alpha$
emission (in addition to the stream passing through the
magnetosphere of the white dwarf and the region of blob collision)
can be expected within the pulsar-like model of AE~Aqr. This
source is associated with the region where the magneto-dipole
waves emitted by the white dwarf are absorbed by the background
material surrounding the system. According to Rees \& Gunn
(\cite{rg74}), the distance to this region, $R_{\rm abs}$, can be
found by equating the pressure of the magneto-dipole radiation,
$p_{\rm md}=L_{\rm sd}/4\pi c R^2$, with the thermal pressure of
the surrounding material, $p_{\rm pl}=(1/2)\rho_{\infty} V_{\rm
s}^2$. The interaction between the waves and the gas leads to the
formation of a shock in which the energy of waves is converted
into the thermal energy of plasma, radiation, and accelerated
particles.

To estimate $R_{\rm abs}$ in AE~Aqr, we have taken into account
that the circumbinary medium of this system is contributed to
mainly by the material ejected due to the propeller action of the
white dwarf. As shown by Wynn et~al. (\cite{wkh97}), this material
flows out within the orbital plane of the system following a
spiral. {\it The position angle of the spiral, however, changes
with the orbital motion of the system}. Therefore, the
distribution of the material surrounding the system has an
azimuthal symmetry. Simulation of the stream-like ejection in the
rotating system has shown that a circumbinary disk-like envelope
with inner radius $R_{\rm cbe} \ga 5\times 10^{11}$\,cm forms
around the system. The mean velocity of the ejected material at
this distance is $V_{\rm b}(R_{\rm cbe}) \sim 100\,{\rm
km\,s^{-1}}$, and therefore the derived scale is comparable to
$P_{\rm orb} V_{\rm b}(R_{\rm cbe})$.

The thickness of the envelope is determined by the thermal
expansion of blobs, and its mean value can be normalized as
$Z_0(R_{\rm cbe}) = 10^{10}\,Z_{10}$\,cm. This allows us to limit
the mean density of the envelope material to
   \be
 \rho_{\rm st} \la 10^{-14}\ l_9^3\ Z_{10}^{-3}\
 \left(\frac{\rho_{\rm b}(r_0)}{10^{-11}{\rm g cm^{-3}}}\right)\
 {\rm g\,cm^{-3}}.
   \ee
Under these conditions the distance at which the magneto-dipole
waves are absorbed by the material of the envelope can be
evaluated as
  \be
R_{\rm abs} \ga 2 \times 10^{12} \rho_{-14}^{-1/2} V_6^{-1}
\left(\frac{L_{\rm sd}}{6\times 10^{33}\,{\rm
erg\,s^{-1}}}\right)^{1/2} {\rm cm}.
 \ee
At this distance the circumbinary envelope occupies about 0.5\% of
the area of a sphere, and the velocity of the material lies within
the interval $\sim 50-150\,{\rm km\,s^{-1}}$. This indicates that
the envelope will contribute to the Doppler tomogram of the system
at low velocities with intensity $\la 5\times 10^{-3} L_{\rm sd}$.

The results of the simulation of the stream trajectories within
the pulsar-like model of AE~Aqr are shown in Figs.\,\ref{tr1}
and~\ref{tr2}. These figures differ in the basic assumption about
the location of the region of the stream disintegration to the
diamagnetic blobs. Namely, in the first run of the calculations
(Fig.\,\ref{tr1}) this region has been placed at the L1 point, and
in the second run we have assumed that the stream disintegration
occurs at the Alfv\'en surface of the primary (Fig.\,\ref{tr2}).
The values of the system parameters (except for $\mu$ and $r_0$)
in all calculations were chosen to be the same as those adopted by
Wynn et~al. (\cite{wkh97}).

In both runs the calculations were made for three different values
of the mass transfer rate: $\dot{M}=10^{16}\,{\rm g\,s^{-1}}$
(panel\,a), $\dot{M}=5 \times 10^{16}\,{\rm g\,s^{-1}}$
(panel\,b), and $\dot{M}=10^{17}\,{\rm g\,s^{-1}}$ (panel\,c). The
value of the parameter $\eta$ was chosen as 0.5. Panel\,(d) shows
the mean picture of the mass transfer on the time scale of the
orbital period of AE~Aqr. This picture is derived by taking a
superposition of states (a), (b), and (c) with equal weights. The
expected mean structure of the H$\alpha$ Doppler tomogram derived
in the first and the second runs are presented at the bottom of
Fig.\,\ref{tr1} and Fig.\,\ref{tr2}, respectively.

One finds the best agreement between the simulated and the
observed tomogram for $<\dot{M}>=5\times 10^{16}\,{\rm
g\,s^{-1}}$, and for $k_0=4.8\times 10^{-7}\,{\rm s^{-1}}$
(assuming the stream to be disintegrated at L1) and $k_0=7.4\times
10^{-7}\,{\rm s^{-1}}$ (assuming the stream to be disintegrated at
$R_{\rm A}$).  Here $<\dot{M}>$ denotes the average mass-transfer
rate on the time scale of the orbital period. Under these
conditions the average value of the Alfv\'en radius of the white
dwarf is $\bar{R_{\rm A}}\simeq 4.9 \times 10^{10}$\,cm. This
allows us to evaluate the average bolometric luminosity of the
stream passing through the magnetosphere as $\bar{L}_{\rm st} \sim
(<\dot{M}>)GM_{\rm wd}/\bar{R_{\rm A}} \simeq 10^{32}\,{\rm
erg\,s^{-1}}$. This value slightly exceeds the luminosity of the
low-velocity source situated beyond the light cylinder of the
white dwarf and is close to the average luminosity of the extended
component of radiation in AE~Aqr evaluated by van Paradijs et~al.
(\cite{pka89}), and Eracleous \& Horne (\cite{eh96}) from the
optical and UV observations. The main features of the derived
tomograms are discussed in the following section.

  \subsection{Comparative analysis of the derived tomograms}

The tomograms calculated within the `pulsar-like' and the
`magnetic propeller' models have a number of similarities. In
particular, neither shows azimuthal symmetry, and the emission is
not centered on the white dwarf but is primarily in the lower-left
quadrant ($V_{\rm x}$, $V_{\rm y}$ both negative). Because of
these properties all of the simulated tomograms resemble the
Doppler tomogram observed from AE~Aqr.

There are, however, several important differences. First, the
upper limit to the velocity of the stream at the closest approach
to the white dwarf within the pulsar-like model is smaller by a
factor of 2 than that within the `magnetic propeller' approach.
This means that the emission associated with the stream passing
through the magnetosphere within the pulsar-like model is produced
at velocities $< 500\,{\rm km\,s^{-1}}$. Since the blobs have
different masses, dispersion of their velocities occurs.
Furthermore, the velocity of blobs at their closest approach to
the primary depends on the mass-transfer rate. Superposition of
these effects leads to a situation in which the contribution of
the stream at $r_0$ appears in the mean Doppler tomogram in the
form of a spread loop, which is centered at ($-250\,{\rm
km\,s^{-1}}; -350\,{\rm km\,s^{-1}})$ and has the size of $|\Delta
V_{\rm x}| \sim 600\,{\rm km\,s^{-1}}$ and $|\Delta V_{\rm y}|\sim
400\,{\rm km\,s^{-1}}$.

The emission at these velocities is present in the observed
H$\alpha$ Doppler tomogram of AE~Aqr. This indicates that blobs
passing through the magnetosphere within the pulsar-like approach
are expected to be hot, and their contribution to the H$\alpha$
emission of the system is significant for all reliable values of
$\dot{M}$. Therefore, the problem of the `missing radiation' from
the loop associated with the blob trajectories (mentioned by Welsh
et~al. \cite{whg98}) does not occur, as one assumes the surface
magnetic field of the white dwarf to be of the order of 50\,MG.

The second feature of the tomogram simulated within the
pulsar-like approach is the significant contribution of the
material situated beyond the light cylinder of the white dwarf.
The radiation of this source is emitted at velocities $\sim
50-150\,{\rm km\,s^{-1}}$ and is powered by the energy of the
magneto-dipole waves. The luminosity of this source is almost
independent of the variations of the mass transfer rate and is
comparable with the luminosity of the stream passing through the
magnetosphere at $\dot{M} \sim 5 \times 10^{16}\,{\rm g\,s^{-1}}$.
The contribution of this low-velocity source to the H$\alpha$
system radiation is seen in the center of the Doppler tomogram as
a spread spot of radius $\sim 150\,{\rm km\,s^{-1}}$.

An additional, intermediate-velocity source of H$\alpha$ emission
is located at a distance of about 1--3 times the binary
separation. This source is associated with the region of possible
blob collisions. Our simulations indicate that in both models
collision of blobs can occur and that the relative velocity of the
colliding blobs is of the order of $V_{\rm b-b} \sim 100\,{\rm
km\,s^{-1}}$. Assuming that all blobs are involved in the
collision process, one can limit the rate of energy release in
this region to $L_{\rm b-b} \la 5 \times 10^{30} \dot{M}_{17}
(V_{\rm b-b}/100\,{\rm km\,s^{-1}})^2\,{\rm erg\,s^{-1}}$. This
indicates that the contribution of this source within the
pulsar-like model can be significant at relatively high
mass-transfer rates but can hardly be recognized when the mass
transfer rate drops below $10^{17}\,{\rm g\,s^{-1}}$.

Finally, our simulations have shown that the structure of the
tomogram within the pulsar-like model is sensitive to variations
in the average mass-transfer rate in the system. As $<\dot{M}>$
decreases, the Alfv\'en radius of the white dwarf becomes larger.
In this case the material moving through the magnetosphere turns
out to be ejected at smaller velocities, and its contribution to
the H$\alpha$ emission of the system decreases. Therefore, the
tomogram, under these conditions, is dominated by the low velocity
component. If, however, $<\dot{M}>$ is large during the period of
observations, the tomogram is dominated by the `high-velocity
spot', which in this case appears in the lower-left quadrant at
velocities $350-500\,{\rm km\,s^{-1}}$. Hence, the observed
night-to-night variations of the tomogram (see Fig.\,10 of Welsh
et~al. \cite{whg98}) can be interpreted within the pulsar-like
model in terms of the variations of the mass-transfer rate in the
system. The range of these variations implies changes in the
efficiency of the propeller action by the white dwarf within the
interval $\bar{L}_{\rm st}/L_{\rm sd} \sim 0.01-0.4$, and
therefore, its contribution to the observed braking of the white
dwarf remains small.

  \section{Discussion}

It is widely believed that the `magnetic propeller' is the only
approach which provides a plausible interpretation of the
H$\alpha$ Doppler tomogram observed in AE~Aqr. Following this
notion, almost all manifestations of the system during the past 5
years have been discussed solely around the hypothesis that the
spin-down of the white dwarf is governed by the propeller
spin-down mechanism.

However, as shown in this paper, the H$\alpha$ Doppler tomogram
expected within the pulsar-like approach also resembles the
observed tomogram. Furthermore, the agreement between the expected
and the observed tomograms within this approach turns out to be
even better than within the `magnetic propeller' model. As
mentioned in Sect.\,3, the basic assumptions adopted within the
pulsar-like approach do not contradict any of the currently
observed properties of the system, but they allow us to invoke the
models developed with respect to the spin-powered pulsars for the
interpretation of properties which AE~Aqr shares with at least
several objects of this class. Therefore, an analysis of the
observed system properties within the pulsar-like approach appears
to be quite reasonable.

The present state of development of both the `magnetic propeller'
and the pulsar-like models is insufficient for recognizing which
of these approaches is more promising. The analysis of this
question is beyond the scope of the present paper. Nevertheless,
to clarify the basic statements of these models we summarize the
assumptions adopted within each of these approaches.

    \subsection{Assumptions adopted within the magnetic propeller
    approach}

The following 6 basic assumptions, currently adopted within the
magnetic propeller model, can be distinguished:

I.~The dipole magnetic moment of the white dwarf is assumed to be
of the order of $10^{32}\,{\rm G\,cm^3}$, i.e. within the expected
range of the magnetic moments of Intermediate Polars.

II.~The spin-down power is assumed to be transferred predominantly
to the kinetic energy of the material ejected from the system due
to propeller action by the white dwarf. As shown in Sect.\,2.1.2,
this assumption implies the mass-transfer rate in the system to be
  \be
\dot{M} \ga 10^{18}\ \left(\frac{r_0}{10^{10}{\rm cm}}\right)
\left(\frac{L_{\rm sd}}{6\times 10^{33}{\rm erg\,s^{-1}}}\right)\
{\rm g\,s^{-1}}.
  \ee

III.~The stream of material transferred from the secondary is
assumed to be strongly inhomogeneous. Actually, this assumption
implies that at least 99.99\% of the material is transferred in
blobs. Indeed, if the mass transfer rate of the homogeneous
component exceeds
 \be\label{dotmhs}
\dot{M}_{\rm hs} \sim 10^{13}\ \eta_{0.37}^{7/2}\ \mu_{32}^2\
M_{0.8}^{-1/2}\ \left(\frac{R}{R_{\rm circ}}\right)^{-7/2}\ {\rm
g\,s^{-1}},
 \ee
the homogeneous flow turns out to be able to reach the
circularization radius and to form a disk around the magnetosphere
of the white dwarf.

It should be noted that the reason for such a strong inhomogeneity
is rather unclear. It might be connected with the magneto-flaring
activity of the normal component or the beam instability in the
region of the L1 point. At the same time, it is unlikely that it
can be explained in terms of the interaction between the stream
and the magnetic field of the white dwarf since at distances $R\gg
R_{\rm A}$ the energy density of the magnetic field is a few
orders of magnitude smaller than the thermal energy of the stream
material. In particular, the solution of Arons \& Lea
(\cite{al80}) derived for the regions $R \sim R_{\rm A}$ is
obviously not applicable in this case.

IV.~The temperature of blobs passing through the magnetosphere is
assumed to be small enough for their contribution to the H$\alpha$
emission of the system to be negligibly small. Otherwise, the
presence of a high velocity loop associated with the trajectories
of blobs interacting with the magnetic field of the white dwarf is
expected (see Fig.\,3 in Wynn et~al. \cite{wkh97}). Such a loop,
however, is not seen in the observed tomogram (for discussion see
Welsh et~al. \cite{whg98}; Welsh \cite{welsh99}).

V.~Almost all the blobs expelled by the white dwarf are involved
in a collision process as their trajectories intersect beyond the
system. Otherwise, the amount of hot material would be
insufficient to explain the observed H$\alpha$ emission. It should
also be noted that for the observed flaring in the system to be
associated with the collision of blobs, the mass transfer rate
should be in excess of (see Sect.\,4.3)
 \be
\dot{M}_{\rm b-b} \ga 1.3 \times 10^{19}\ L_{33}\
\left(\frac{V_{\rm b-b}}{123{\rm km\,s^{-1}}}\right)^{-2} {\rm
g\,s^{-1}},
 \ee
where $L_{33}$ is the luminosity of the flaring component
expressed in units of $10^{33}\,{\rm erg\,s^{-1}}$ (see van
Paradijs et~al. \cite{pka89}; Beskrovnaya et~al. \cite{bibs96}),
and the value of $V_{\rm b-b}$ is normalized to the rising
velocity of flares recently derived by Skidmore et~al.
(\cite{soh03}) from high-time-resolution spectroscopy of the
system. Otherwise, the energy release in the region of blobs
collision turns out to be smaller than the luminosity of strong
flares observed from the system.

VI.~One has to assume that the blobs cool down very slowly (on a
time scale of a few hours) or, for some reason, are heated again
at larger distances from the system. Otherwise, the origin of the
low-velocity ($\la 100\,{\rm km\,s^{-1}}$) component of the system
H$\alpha$ emission, which is seen on the observed tomogram,
becomes rather unclear (see e.g. Welsh \cite{welsh99}; Pearson
et~al. \cite{phs03}).

Within these assumptions a good agreement between the observed
H$\alpha$ Doppler tomogram and the tomogram simulated within the
`magnetic propeller' model can be achieved.

   \subsection{Assumptions adopted within the pulsar-like white
   dwarf approach}

The basic assumptions of the pulsar-like model of AE~Aqr are as
follows:

I.~The dipole magnetic moment of the white dwarf is assumed to be
as high as $\mu \simeq 1.4 \times 10^{34}\,{\rm G\,cm^3}$. As
shown in Sect.\,3, this assumption implies a rather complicated
scenario for the system evolution, which invokes processes poorly
investigated so far.

II.~The stream of material transferred from the secondary is
assumed to disintegrate into large diamagnetic blobs as it impacts
onto the Alfv\'en surface of the primary.

III.~It is assumed that the mass-transfer rate in the system
varies by a factor of a few on the time scale of the orbital
period. Due to these variations the structure of the tomogram
changes from night to night.

IV.~We have also assumed that the stream does not strongly
interact with relativistic particles accelerated in the potential
gap situated in the vicinity of the white dwarf surface. This
implies that the opening angle of the beam of particles
accelerated in the magnetic pole regions is $\theta \la
18^{\degr}$ (here, the value of $\beta$ is adopted as $\la
77^{\degr}$, see Eracleous et~al. \cite{erac94}). If,
nevertheless, there is interaction between the stream and the
accelerated particles, the energy balance in the material of the
stream should be re-calculated.

At the same time, none of the assumptions adopted within the
`magnetic propeller' approach is necessary in the frame of the
pulsar-like model of AE~Aqr. In particular, there is no strong
lower limit to the mass-transfer rate, diskless mass transfer is
expected independently of whether the stream inflowing through the
L1 point is homogeneous or inhomogeneous, the blobs passing the
accelerating region are expected to be hot, and the low-velocity
source is associated with the interaction between the
magneto-dipole waves and the plasma expelled by the propeller
action of the white dwarf.

  \section{Conclusions}

We have shown that the H$\alpha$ Doppler tomogram simulated within
the pulsar-like white dwarf model of AE~Aqr is similar to the
observed tomogram in several important aspects. Namely, the
emission is not centered on the white dwarf, it does not show
azimuthal symmetry, and the strongest emission occurs primarily in
the lower-left quadrant at velocities $\la 500\,{\rm km\,s^{-1}}$.

At least three sources of the H$\alpha$ emission can be
distinguished within the considered approach: (1)~the emission of
the stream passing through the magnetosphere of the white dwarf
(the high-velocity component: $350-500\,{\rm km\,s^{-1}}$),
(2)~the region of blob collisions (the intermediate velocity
component: $200-300\,{\rm km\,s^{-1}}$), and (3)~the region of
interaction between the stream and the magneto-dipole radiation of
the white dwarf (the low-velocity component: $\la 150\,{\rm
km\,s^{-1}}$). The relative contributions of these components to
the system emission depend on the mass-transfer rate, and
therefore the structure of the tomogram is expected to vary as the
rate of mass transfer from the normal companion into the Roche
lobe of the white dwarf changes.

The best agreement between the simulated and the observed
tomograms was found assuming that the mass-transfer rate varies on
a time scale of a few hours in the interval $\dot{M} \sim 10^{16}
- 10^{17}\,{\rm g\,s^{-1}}$ with the nightly mean value $<\dot{M}>
\simeq 5\times 10^{16}\,{\rm g\,s^{-1}}$. In this case the
efficiency of the propeller action by the white dwarf is limited
to $0.01 \la L_{\rm st}/L_{\rm sd} \la 0.4$. This means that the
contribution of the propeller spin-down mechanism to the observed
braking of the white dwarf under the conditions of interest is
small, and hence, the spin-down of the white dwarf (which is
assumed to be governed by the pulsar-like spin-down mechanism) is
expected to be stable independently of variations of $\dot{M}$.

\begin{acknowledgements}
   We thank Dr. Chul-Sung Choi and an anonymous referee for very
   careful reading of the manuscript and useful comments. Nazar
   Ikhsanov acknowledges the support of the Alexander von Humboldt
   Foundation within the ``Long-term Cooperation'' Program. Vitaly
   Neustroev acknowledges the support of IRCSET under their basic
   research programme and the support of the HEA funded CosmoGrid
   project. The work was partly supported by the Russian Foundation
   of Basic Research under the grant 03-02-17223a and the State
   Scientific and Technical Program ``Astronomy''.
\end{acknowledgements}


\begin{thebibliography}{}
\bibitem[1980]{al80}
  Arons, J., Lea, S.M. 1980, ApJ, 235, 1016
\bibitem[1988]{bdch88}
  Bastian, T.S., Dulk, G.A., Chanmugam, G. 1988, ApJ, 324, 431
\bibitem[1997]{bt97}
  Becker, W., Tr\"umper, J. 1997, A\&A, 326, 682
\bibitem[1996]{bibs96}
  Beskrovnaya, N.G., Ikhsanov, N.R., Bruch, A., Shakhovskoy, N.M. 1996,
        A\&A, 307, 840
\bibitem[1992]{bbch92}
  Bowden, C.C.G., Bradbury, S.M., Chadwick P.M., et~al. 1992,
  Astropartical Physics, 1, 47
\bibitem[1991]{bruch91}
  Bruch, A. 1991, A\&A, 251, 59
\bibitem[1992]{chanm92}
  Chanmugam, G. 1992, ARA\&A, 30, 143
\bibitem[1987]{cmt87}
  Chanmugam, G., Meenakshi, R., Tohline, J.E. 1987, ApJ, 319, 188
\bibitem[1981]{chw81}
  Chincarini, G., Walker, M.F. 1981, A\&A, 104, 24
\bibitem[1999]{cda99}
  Choi, C-S., Dotani, T., Agrawal, P.C. 1999, ApJ, 525, 399
\bibitem[2000]{cy00}
  Choi, C.-S., Yi, I. 2000, ApJ, 538, 862
\bibitem[1995]{co95}
  Clayton, K.L., Osborne, J.P. 1995, in ``Magnetic Cataclysmic Variables'',
     eds. D.~Buckley and B.~Warner, ASP Conference Series 85, 379
\bibitem[1990]{crop90}
  Cropper, M. 1990, Space Sci. Rev., 54, 195
\bibitem[1994]{jmor94}
  de~Jager, O.C., Meintjes, P.J., O'Donoghue, D., Robinson, E.L. 1994,
     MNRAS, 267, 577
\bibitem[1996]{eh96}
  Eracleous, M., Horne, K. 1996, ApJ, 471, 427
\bibitem[1994]{erac94}
  Eracleous, M., Horne, K., Robinson, E.L., et~al. 1994, ApJ, 433, 313
\bibitem[1985]{fkr85}
  Frank, J.F., King, A.R., Raine, D.J. 1985, ``Accretion power in
     Astrophysics'', Cambridge University Press, Cambridge, p.\,60
\bibitem[1997]{friedjung97}
  Friedjung, M. 1997, New Astron., 2, 319
\bibitem[1969]{gj69}
  Goldreich, P., Julian, W.H. 1969, ApJ, 157, 869
\bibitem[1986]{hkl86}
  Hameury, J.-M., King, A.R., Lasota, J.-P. 1986, MNRAS, 218, 695
\bibitem[1995]{har95}
  Hartmann, D.H. 1995, A\&A Rev., 6, 225
\bibitem[1997]{i97}
  Ikhsanov, N.R. 1997, A\&A, 325, 1045
\bibitem[1998]{i98}
  Ikhsanov, N.R. 1998, A\&A, 338, 521
\bibitem[1999]{i99}
  Ikhsanov, N.R. 1999, A\&A, 347, 915
\bibitem[2001]{i01}
  Ikhsanov, N.R. 2001, A\&A, 374, 1030
\bibitem[2002]{ijb02}
  Ikhsanov, N.R., Jordan, S., Beskrovnaya, N.G. 2002, A\&A, 385,
  152
\bibitem[2002]{ib02}
   Ikhsanov, N.R., Beskrovnaya, N.G. 2002, ApJ, 576, L57
\bibitem[2001]{jord01}
  Jordan, S. 2001, in ``White Dwarfs'', eds. J.L.\,Provencal,
  et~al., ASP Conference Series, 226, 269
\bibitem[1998]{kr98}
  Klu\'zniak, W., Ruderman, M. 1998, ApJ, 505, L113
\bibitem[1998]{lbc98}
  Lang, M.J., Buckley, J.H., Carter-Lewis, D.A., et~al. 1998,
  Astroparticle Physics, 9, 203
\bibitem[1977]{mt77}
  Manchester, R.N., Taylor, J.H. 1977, ``Pulsars'', San Francisco:
     Freeman
\bibitem[2002]{m02}
  Meintjes, P.J. 2002, MNRAS, 366, 265
\bibitem[1994]{mjr94}
   Meintjes, P.J., de\,Jager, O.C., Raubenheimer, B.C., et~al. 1994,
   ApJ, 434, 292
\bibitem[2000]{mdj00}
   Meintjes, P.J., de~Jager, O.C. 2000, MNRAS, 311, 611
\bibitem[2003]{mv03}
   Meintjes, P.J., Venter, L.A. 2003, MNRAS, 341, 891
\bibitem[1991]{m91}
  Michel, F.C. 1991, ``Theory of Neutron Star Magnetospheres'',
     University of Chicago Press
\bibitem[1968]{pac68}
  Pacini, F. 1968, Nature, 219, 145
\bibitem[1990]{pac90}
  Paczy\'nski, B. 1990, ApJ, 365, L9
\bibitem[1979]{p79}
  Patterson, J. 1979, ApJ, 234, 978
\bibitem[1980]{pbchr80}
  Patterson, J., Branch, D., Chincarini, G., Robinson, E.L. 1980,
  ApJ, 240, L133
\bibitem[2003]{phs03}
  Pearson, K.J., Horne, K., Skidmore, W. 2003, MNRAS, 338, 1067
\bibitem[1974]{rg74}
  Rees, M.J., Gunn, J.E. 1974, MNRAS, 167, 1
\bibitem[1994]{rb94}
   Reinsch, K., Beuermann, K. 1994, A\&A, 282, 493
\bibitem[1995]{rbht95}
   Reinsch, K., Beuermann, K., Hanusch, H., Thomas, H.-C. 1995, in
     ``Magnetic Cataclysmic Variables'', eds. D.~Buckley and B.~Warner,
     ASP Conference Series 85, 115
\bibitem[2002]{skkwz02}
   Schenker, K., King, A. R., Kolb, U., et~al. 2002, MNRAS, 377,
   1105
\bibitem[2003]{soh03}
  Skidmore, W., O'Brien, K., Horne, K., et~al. 2003, MNRAS, 338, 1057
\bibitem[1999]{s99}
  Spruit, H. 1999, A\&A, 341, L1
\bibitem[1992]{ssb92}
  Stockman, H.S., Schmidt, G.D., Berriman, G., et~al. 1992, ApJ, 401, 628
\bibitem[1996]{tom96}
  Tompson, D.J. 1996, in ``Pulsars: Problems \& Progress'',
     eds. S. Johnston, M.A. Walker, and M. Bailes, ASP Conference
     Series, 105, 307
\bibitem[1989]{pka89}
  van Paradijs, J., Kraakman, H., van Amerongen, S. 1989, A\&AS, 79,
     205
\bibitem[1995]{w95}
  Warner, B. 1995, ``Cataclysmic Variable Stars'', Cambridge
     University Press.
\bibitem[1999]{welsh99}
  Welsh, W.F. 1999, ``Annapolis Workshop on Magnetic Cataclysmic Variables'',
     eds. C.~Hellier and K.~Mukai, ASP Conference Series 157, 357
\bibitem[1995]{whg95}
  Welsh, W.F., Horne, K., Gomer, R. 1995, MNRAS, 275, 649
\bibitem[1998]{whg98}
  Welsh, W.F., Horne, K., Gomer, R. 1998, MNRAS, 298, 285
\bibitem[1995]{wk95}
  Wynn, G.A., King, A.R. 1995, MNRAS, 275, 9
\bibitem[1997]{wkh97}
  Wynn, G.A., King, A.R., Horne, K. 1997, MNRAS, 286, 436
\end{thebibliography}
\end{document}